# What is a Fog Node?
# A Tutorial on Current Concepts towards a Common Definition

Eva Marín Tordera[*], Xavi Masip-Bruin[*], Jordi García-Almiñana[*], Admela Jukan[§]
Guang-Jie Ren[‡], Jiafeng Zhu[**], Josep Farré[†],

[*]Universitat Politècnica de Catalunya-Advanced Network Architectures Lab, UPC-CRAAX, Spain
{eva, xmasip, jordig}@ac.upc.edu

[§] Technische Universität Braunschweig, Germany
a.jukan@tu-bs.de

[‡]IBM, Almaden Research Center, USA, gren@us.ibm.com

[**]Huawei, Santa Clara, USA, jiafeng.zhu@huawei.com

[†]Neàpolis Center, Spain, josepft@neapolis.cat

**Abstract:**

Fog computing has emerged as a promising technology that can bring the cloud applications closer to the physical IoT devices at the network edge. While it is widely known what cloud computing is, and how data centers can build the cloud infrastructure and how applications can make use of this infrastructure, there is no common picture on what fog computing and a fog node, as its main building block, really is. One of the first attempts to define a fog node was made by Cisco, qualifying a fog computing system as a "mini-cloud," located at the edge of the network and implemented through a variety of edge devices, interconnected by a variety, mostly wireless, communication technologies. Thus, a fog node would be the infrastructure implementing the said mini-cloud. Other proposals have their own definition of what a fog node is, usually in relation to a specific edge device, a specific use case or an application. In this paper, we first survey the state of the art in technologies for fog computing nodes as building blocks of fog computing, paying special attention to the contributions that analyze the role edge devices play in the fog node definition. We summarize and compare the concepts, lessons learned from their implementation, and show how a conceptual framework is emerging towards a unifying fog node definition. We focus on core functionalities of a fog node as well as in the accompanying opportunities and challenges towards their practical realization in the near future.

**Keywords:** fog computing, fog node, IoT, edge devices, edge computing

## 1. Introduction: Moving cloud to the edge

Cloud computing has become an essential information technology power horse, commonly used by a myriad of applications, and valued by users to seamlessly run business, entertainment and social network applications at remote data center premises. The IT outsourcing feature of the cloud is not only bringing value added services, but also lowering expectations on the ability of edge devices to process the applications locally. The recent proliferation of Internet of Things (IoT)-related services, including eHealth [1], smart cities [2], smart transportation systems [3] and industrial scenarios [4], to name a few, are however challenging the performance of cloud computing, mostly for the reasons of unpredictable and often high communication latency, privacy gaps and related traffic loads of networks connecting cloud computing to end-users. To address some of these limitations of cloud computing, the research community has recently proposed the concept of *Fog Computing*, aiming at bringing cloud service features closer to what is referred to as "Things," including sensors, embedded systems, mobile phones, cars, etc.

Fog computing was initially proposed in the area of IoT to help execute applications and services. The work by Al-Fuqaha [5], et al, surveyed IoT concepts with fog computing to deploy IoT applications, such as, location, distribution, scalability, density of devices, and mobility support. The first more formal definition of fog computing, by Bonomi et all in [6], stated that *'Fog computing is a highly virtualized platform that provides compute, storage and networking services between end devices and traditional Cloud computing Data Centers, typically, but not exclusively located at the edge of the network"*. Similar definition can be found in [7] stating that *'Fog computing is proposed to enable computing directly at the edge of the network, which can deliver new applications and services especially for the future of Internet'*.

In fact, a number of surveys focused on fog computing exists, see [8], [9] and [10], aiming at revisiting fog computing concepts, thus defining what fog computing is, its challenges, possible applications as well as scenarios where fog computing may undoubtedly contribute to. Unlike the set of existing contributions surveying what fog computing is, this paper is not intended to revisit fog computing as a novel cloud paradigm, but with special emphasis on fog computing infrastructure deployment, contributing to the common understanding and a well-understood definition of a fog node as its main endeavor. Aligned to that objective, this paper extends the traditional scope of fog computing surveys scope by proposing novel resource organization concepts for a fog node leveraging abstractions and virtualizations of heterogeneous physical edge devices as the key pillar to accommodate physical edge devices heterogeneity.

It must be highlighted that the formal concept of fog computing is not disruptive. Since its inception, the main fog computing model has been perceived as what is known as edge computing, including cloudlets [11], Mobile Edge Computing [12], Intelligent Transport Systems Clouds (ITS-Clouds) [13] and VANET (Vehicular Ad Hoc NETworks)-Clouds [14]. The overarching idea in all these concepts is to make it possible to run applications based on location and closer to the user, on virtualized hardware devices, as we have seen in mobile clouds, cloudlets, ITS-clouds, etc. Related to it are also efforts in the so-called Mobile Grid Computing (see [15][16][17]). The Mobile Cloud Computing [18] paradigm (MCC) is also close to the edge computing concept, as it aims at providing solutions to guarantee an efficient offloading of applications and services from mobile devices to remote resource providers –cloud, fog or cloudlets [19]. Intimately related to MCC, in Mobile Device Clouds (MDC) mobile devices offload their tasks to local clouds built by grouping neighboring edges (see for example [20][21][22]). Similar ideas can be found in Content Delivery Networks (CDN) [23]. In CDNs, cache servers are deployed at the edge of the network to reduce the latency when downloading content from remote sites. Table 1 illustrates some of the relevant edge computing proposals as they appear in the various categories and flavors. Although significant differences may be highlighted between these concepts, they all essentially propose the use of proximate computational resources rather than remote resources in datacenters. Some differences, we believe, stem from the research communities addressing them. For example, cloudlets come from the cloud computing research community while fog computing comes from the area of networking. The similarity/differences between these concepts are summarized in [8], stating that *"some other concepts, not declared as 'fog computing', might fall under the same 'umbrella' e.g., cloudlets.'*.

**Table 1.** *Edge computing categories*

| Techology | References | Including mobile edge devices |
|---|---|---|
| **Fog computing** | [1][2][3][4] [6][9] [20] [31] [24] [27][28][29][30] [26][32][34] [33][46][47][48] | Only when fog nodes include underlying edge devices |
| **Cloudets** | [11] [19] [56] | No |
| **Mobile Edge Computing** | [12] | Yes, by definition |
| **ITS and VANET clouds** | [13] [14] [35] [59] | Yes, vehicles |
| **Mobile Device Cloud** | [20][21][22] [57] [58] | Yes, the mobile user offloads to cloud and to other edge mobile devices |
| **Content Delivery Networks** | [23] | No |
| **Grid proposals including edge devices (such as Mobile Grid Computing)** | [38][15][16][17] | Yes |

If it can be agreed upon that fog is part of a broader concept of *edge computing*, the open question remains whether there can be a clear and well-accepted definition of the basic functional and conceptual entity in fog computing, i.e., the fog node. So far, a fog node was considered as a physical that implements fog computing. –i.e., "what"–, as read in [7] '*In fog computing, facilities or infrastructures that can provide resources for services at the edge of the network are called fog nodes.*', or in [24], '*a fog node is the physical device where fog computing is deployed*'. However, there is no consensus on "how" and "where" a fog node is implemented.

Starting with the seminal paper by Bonomi [6], the fog node concept is not defined, hence it yet opens the possibilities for various interpretations. In one interpretation, fog computing consists of a set of embedded systems, smart phones, actuators and sensors forming the Smart Things Network. As such, the fog computing system may be seen as a data producer/consumer layer, i.e., raw data from sensors, or pre-processed information from embedded systems generated and forwarded to be processed elsewhere. In another interpretation, fog can be seen as a computing layer. Indeed, some of the systems and sensors building this layer have CPU and storage capacities, which would be used not only to process their own data but also to process external requests. We also analyze the attempts to craft a more concrete definition for a fog node as in [24]. While the report in [24] does not conclude with a single fog node implementation strategy, it proposes a variety of devices as candidates, including routers, switches, wireless access points, video surveillance cameras, and Cisco Unified Computing System (UCS) servers. A common characteristic for all these devices to become potential fog nodes, is that they all embed computing, storage and networking capabilities, essential to ease the execution of IoT applications [25].

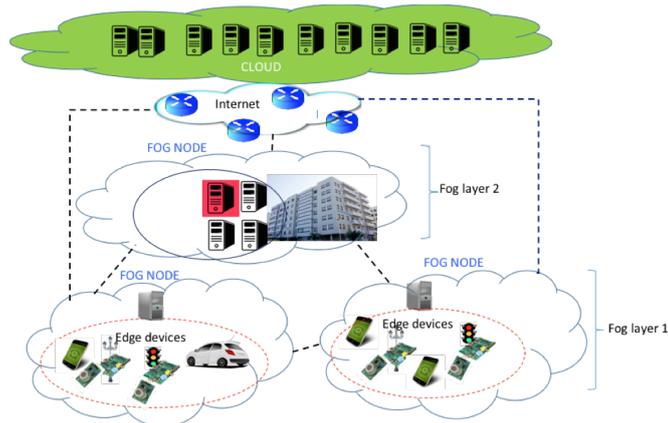

**Figure 1.** *Fog-to-cloud architecture (F2C)*

From our perspective, the most interesting aspect of a fog node definition is that we effectively need a system that can, on the one hand control a specific set of edge devices, while on the other, access to clouds. Following this rationale, this paper also frames some thoughts on the coordinated Fog to Cloud (F2C) management architecture in [26], and the role of fog node concept within. Let us assume a layered architecture connecting a set of devices and capabilities with a stack of resources, creating a cloud layer, and a few different hierarchical fog layers vertically and horizontally distributed, akin to the OSI layer in networking. Figure 1 depicts a possible F2C scenario including one cloud layer and two fog layers: fog layer 1 directly connected to the edge devices – mobile phones, sensors, small processing boards, etc.–, and fog layer 2 standing for an intermediate computing capacity layer, between fog layer 1 and cloud.

In this F2C hierarchical scenario, we envision fog nodes at the lowest fog layer, directly connected to edge devices, devoted to aggregate and control the edge devices capacities – storage, sensing, computing and network. Towards this vision, there is a myriad of concepts, approaches and ideas that can be evaluated on their suitability to help define fog nodes in future systems. The main goal of this paper is systematically address issues relevant to providing a common definition for a fog node.

When comparing with cloud computing, there is no need for a formal definition for a cloud node mainly because cloud computing is per se centralized. However, when moving to fog computing, its distributed nature and heterogeneity makes, in our opinion, the reason for the suggested fog node definition.

This paper is structured into two main parts. The first part surveys the existing fog node related concepts, including functional and conceptual approaches to define a fog node, and its relationship with edge devices (Sections 2 and 3). The second part opens up the discussion on what a fog node could be, discussing open issues and challenges, thus laying the foundations for a common fog node definition (Sections 4 and 5). Finally, Section 6 concludes the paper.

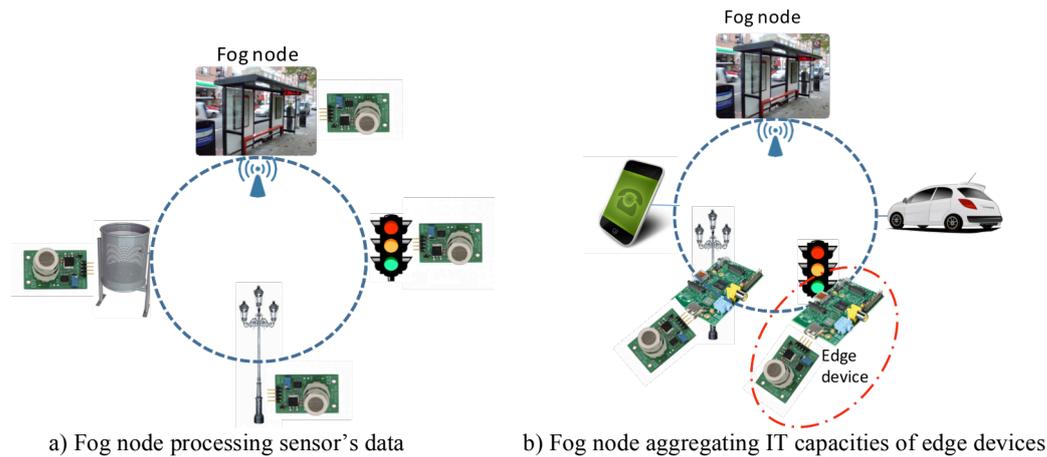

| a) Fog node processing sensor's data | b) Fog node aggregating IT capacities of edge devices |

**Figure 2.** *Two categories of fog nodes based on the type of edge device*

## 2. Revisiting fog node concepts: How and Where

We start the survey contribution of this paper by categorizing fog nodes into two categories (the "how"), differing in the role that edge devices play –i.e., characteristics and functionalities–, including: 1) fog nodes as mini-clouds with ("dumb") edge devices acting as data producers/consumers, and; 2) fog nodes as mini-clouds with ("smart") edge devices enriched with IT capacities. It should be noted that our main goal is to classify edge devices depending on their characteristics and functionalities. To that end we learn from the edge devices description details provided by the papers referenced. When these details are not available, we infer the said characteristics from apps and services to be developed at that device, their location, etc.

There are several ideas in the recent literature falling into the first category, including the industry-led proposals [24] and [25], the Smart Gateway proposed in [27], the eHealth services in [28], the micro data centers proposed in [29], or the proposal of fog nodes serving as caches in Information Centric Networking found in [30]. The second category includes the early contribution on fog computing in [31], proposing a three-layered architecture, consisting of cloud data centres, fog nodes at the edge of the network and devices as end points. Closely related contributions in the second category are also in [2] and [32], leveraging the fact that applications are distributed across different layers and all assuming that edge devices collect and process the raw data gathered from sensors.

Let now us illustrate the two said categories through the instructive example shown in Figure 2. Suppose a smart city platform includes several bus stops equipped with fog nodes designed as mini-clouds that comprise at least one server with processing capabilities. Assume first that bus stops are fog nodes falling into the first category, i.e., edge devices are only data producers/consumers, what means that each bus stop can control a city area with different sensors and actuators (Figure 2.a) For example, sensors would measure the CO and $CO_2$ levels in the city and forward this data for processing to the fog node installed in the bus stop. The fog node in the bus stop will process the data collected from the different sensors in its coverage area, possibly merged with either other data gathered from other sensors (e.g., temperature, number of detected vehicles, etc.) or even with information downloaded from existing and data-related repositories at cloud. For instance, the goal of this processing may be to issue a warning to the city management, aimed at limiting the number of cars in that area of the city.

Let us now analyze the second category in the same scenario, as illustrated in Figure 2.b. Here, the edge devices are now enriched not only with sensing but also computing capacities, including a sensor connected to a compute board, a mobile phone and a car (we assume that the car and the mobile phone also include a temperature sensor). The data processing is now possible by the edge device itself, hence producing information –pre-processed data–, to be forwarded to the mini-cloud within the fog node.

**Table 2.** *Fog node location*

| Gateways | [1][27][28][29] |
|---|---|
| Intermediate compute nodes | [2] [3] [6][31] |
| Network elements such as routers | [3][4][6][31][24][30][32] |

Another key aspect in the definition of a fog node, as already highlighted in Section 1, is also "where" fog nodes are located. Table 2 shows most common fog node locations. Some of the revisited contributions propose to locate fog nodes in highly capable devices, such as routers or smart gateways. In such a group we may include contributions in [1], [27], [28] and [29] proposing the use of gateways for deploying fog computing in different scenarios (for example [1] and [28] both focus on eHealth). More in detail, in [1] the gateway is proposed to act as an intermediate point between sensors connected to the patient and the local switch/Internet, receiving data from the sensors, running some protocol conversion, and feeding the upper layer with services, such as data aggregation, filtering and dimensionality reduction. Authors in [28] propose to enrich the traditional gateway functionalities, with the capacity to pre-process data coming from an ECG device (Electrocardiogram). In a different scenario, authors in [27] propose a smart gateway to connect IoT producing audio or video data to the cloud. The proposed smart gateway is augmented with the capacity to process the data to be forwarded to the cloud via Internet. In [29], a smart gateway is proposed to implement the so-called fog micro-data center supporting functions of resource estimation and management.

Works [3], [6] and [31], proposed by Bonomi et al., established the foundations of fog computing as an intermediary computing layer between cloud resources and edge devices, designed in an open fashion such there is no dependency from specific devices. Authors in [2] propose the use of three fog computing layers for big data analysis in smart cities. The first fog layer, called intermediate computing nodes, consists of computers with intermediate computer power. The second fog layer, called edge computing nodes, is built by small computing nodes (e.g., mobile phones). Finally, the third layer, called the lowest fog layer, consists of sensors with sensing capacities only.

In a different application [4], fog computing was applied in the industrial environment. In that paper, fog computing is implemented in Cisco edge routers, as it was first proposed by Cisco in [24]. Authors in [30] propose to deploy fog computing in routers at the edge of the network to implement ICN (Information Centric networking). In [32] authors also propose a three layered architecture, known as Cloud, Fog and Dew. In this structure, the Dew layer refers to the edge devices (e.g., sensors or cameras) and the Fog layer is implemented at devices at the edge of the network (e.g., network routers), and is responsible for providing compute, storage and application services closer to edge devices producing the data.

The two fog node categories mentioned above, i.e., the fog nodes with "dumb" edge devices that can only produce data, and fog nodes with "smart" edge devices preprocessing data, however, do not paint the full picture for the envisioned scenarios to come in the near future. Innovative highly demanding services (for example a medical emergency service [26]) may require additional processing and storage capacity from a richer set of resources not included in the two categories above. Different solutions are possible depending on the services envisioned and the resources management strategy. For example, in emergency scenarios (natural disasters, etc.), computing capabilities can be acquired on demand, from volunteer sources, such as cars parked nearby, or individuals offering their smart phone resources to the emergency personnel if they happen to be close by. In [15], an interesting example was given of sharing smart phone computation resources only when phones are connected to the grid. Thus, adding processing capacities to edge devices seems to be not enough to handle such highly demanding scenarios, therefore further concepts must be developed. Such advanced concepts are innovative and rather interesting, but likely to increase the complexity of the overall system, requiring not only common but also standardized abstractions of the heterogeneous edge devices. While some previous work proposed abstraction methods for sensors, we argue that not only the sensor devices are to be provided with standardized abstraction, but also other types of edge devices in a shared resource pool.

### 3. Meeting new challenges: Further edge devices categorization

As previously introduced, the concept of a fog node is essentially based on the characteristics and features of edge devices, turning into two categories. In one category, "dumb" edge devices are producing data (such as sensors), or acting as actuators. In the second category, "smart" edge devices include various compute, storage and networking capabilities.

*3.1 Further smart edge devices categorization*

The focus of this section is mainly on the second category, i.e., smart edge devices. This is motivated by what we believe is the vision of a near future, where different IoT devices with IT/sensing capacities can be used to create a large scale *grid* (or more than one *grid*) by means of novel sharing or collaborative policies. In a smart city scenario, for example, what is today a cloud service may be an entity able to request IT/sensing resources from the *grid*. A salient feature in this collaborative scenario is the ability of edge devices to offering their capacities, be it sensing and/or processing.

In the category of the said smart edge devices, further differentiation can be made between specialized devices, which only process the data from the sensors, Fig 3.a, they are connected to, and general-purpose devices offering their resources for sharing, Fig 3.b, with various degrees of IT capacities, from smart phones to multi-platform

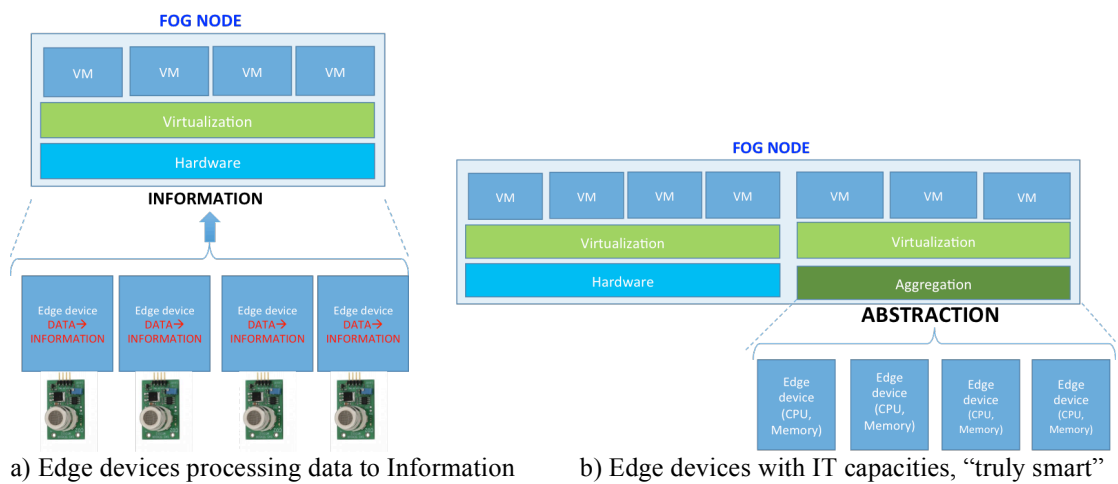

a) Edge devices processing data to Information    b) Edge devices with IT capacities, "truly smart"

**Figure 3.** *Edge devices function*

management, such as clusters, grids, and ITS clouds. We refer to these edge devices as "truly smart". Figure 3.a and Figure 3.b illustrate different capabilities of smart edge devices, including the need for IT capacity abstraction (see Figure 3.b).

Summarizing the role of edge devices in fog computing, we recall three basic categories of edge devices: i) "dumb" devices as mere data producers/consumers; ii) "smart" edge devices with the capacity to process (only) their own data, and; iii) "truly smart" edge devices offering their IT capacity to run distributed applications. Next subsection revisits current work describing edge devices functionalities.

*3.2 Related work reviewed*

We summarize in Table 3 the prior art in fog computing according to the set of functionalities embedded in the edge devices. The first three rows correspond to fog proposals with edge devices in the role of: i) mere data producers ("dumb"), ii) edge devices processing their own local raw data from connected sensors ("smart"), and iii) edge devices offering their IT capacity to execute external services ("truly smart"). Recall that the classification about the role of edge devices is obtained from the details provided by the papers reviewed, and when these details are not explicit enough, we inferred this information from overall context in the paper.

Table 3. *Edge devices functionality*

| | |
|---|---|
| **Fog with edge devices as mere data producer/consumer** | [1][3][6] [27][28][29][30][58] |
| **Fog with edge devices processing local data** | [2][31][32] |
| **Fog with edge devices offering their IT capacity** | [20] [26] [34] [9][33] |
| **Edge devices offering their IT capacity (distributed computing)** | [14] [38] [15][16][17] [35] [39] |
| **Offloading to edge devices** | [21][22] |

In the fourth row of the table we list proposals dealing with distributed computing. Though not directly linked to fog computing, proposals into the fourth row are listed intended to both, comparison purposes and to set a complete spectrum of options. In this fourth row we summarize some of the existing works categorized as distributed computing including edge devices, such as VANET clouds, Jungle computing, mobile grid computing, volunteer computing, where edge devices may be cars, mobile phones, etc. All proposals listed in these four rows require a triggering feature to run the application. In other words, all these options include some sort of management system and/or resource coordinator that allocates tasks to resources.

The fifth row on the other hand, also considers the distributed execution using edge devices (mobile phones), whereby, unlike the papers listed in the forth row, the application here is initiated at an edge device and part of the execution is offloaded to other edge devices.

Figure 3.a illustrates what is closest to the proposal in [31], where the concept was discussed that edge devices can run either an application, or a small portion of an application, whereby the output information from the edge device is not just the raw data, but rather a piece of elaborated information obtained through pre-processing in the edge device. In other words, edge devices process data collected from sensors/actuators they are connected to. The pre-processing is a highly beneficial feature as it reduces the amount of data sent to the fog node throughout the network, while offloading the pre-processing to the edge devices. This was in fact studied in [32], where the idea was to endow edge devices with collecting/generating and pre-processing capacities, turning raw data into information, which is propagated to higher levels, be it fog and/or cloud. Paper [2], on the other hand, proposes a hierarchical, layered-based architecture for big data Analysis in Smart Cities, whereby Layer 1 is Cloud and Layer 2, Layer 3 and Layer 4 are considered three fog layers, as already described in section 2. More in detail, smart edge devices in layer 3 can collect, aggregate, identify potential threat patterns – with applications of machine learning algorithms–, and finally convert the sensors' collected raw data into information.

When the computing capacity is embedded in edge devices only to process local area sensor's data, as previously described, there is no need to either abstract or aggregate their capacity such it can be offered to another process of resource discovery, see Figure 3.a. This is not the case with the third category of edge devices, "truly smart", where the role of edge device further extends towards richer IT capacities, as shown in Figure 3.b, hence driving the need for resource aggregation and abstraction.

Figure 3.b illustrates the features of aggregation and abstraction of edge devices in Figure 2.b. Here, the board attached to the sensor can also run applications not necessarily related to the data collected from the same sensor connected to it. In a similar fashion, the car or the mobile phone, can also share their computational power to process service requests coming from an external service management system. As the computational power of edge devices is offered to run services in a distributed fashion, the challenge is to integrate such a distributed set of edge devices with a cloud computing system, another fog computing system or with a new edge device. In this context, paper [9] analyses the definition and role of fog computing, specifically discuss the edge-cloud –referred in this paper as mini-cloud–, and virtualized sensor networks. In their approach, applications are divided into *droplets*, tiny pieces of code running at edge devices, thus removing the unnecessary upload of data to central servers. Another proposal, called Mobile Fog [33], proposes a programming model to run hierarchically distributed applications according to their workload at cloud, fog and edge devices. Here, the fog nodes were defined as physical devices located inside the network infrastructure, and connected to mobile edge devices to include smart phones, vehicles, etc. The paper proposes an application, to be executed by invoking a specific function – called *connect_fog()* –, enabling the edge device to set a fog process connecting to the global Mobile Fog process running on a fog node.

Another related example can be found in [34], where it was proposed to use mobile phones to perform data analytics in IoT applications, whereby users offered any available resources based on some access policies and resource sharing principles. The main innovation here was to match the partitioning of the application data according to the capacity of the existing resources at the participating edge devices. In [20], a service is divided into different tasks and substasks, hence enabling potential offloading towards either neighboring edge devices or to the cloud.

Similar proposals to distribute application execution have been made also elsewhere. VANET (Vehicular Ad Hoc NETworks) Cloud proposals in [14] and [35] have also considered sharing of resources through edge devices, in this case vehicles, as compute entities. In [35] it was proposed that in a fleet of cars, either one vehicle is appointed as the cloud controller, or all vehicles can act as the interface to the cloud networks. Similar proposals in the area of Mobile Device Clouds (MDC) can be found in [21][22].

Finally, the application execution in different clouds has been referred to as multi-platforms (clusters, grids and cloud) -not included in Table 3-, stemming from the areas of high performance computing and parallel programming. Among a myriad of previous work proposing usage of using multiple platforms, of particular interest are those that combine cloud with other type of resources (grid and clusters) and the corresponding multi-platform resource allocation methods. For instance, applications are executed in a distributed fashion in different clouds [36], in a combination of cloud and grid resources [37], or in a combination of heterogeneous, hierarchical distributed and high performance resources, such as in *Jungle Computing* [38]. *Jungle Computing* represents the extreme case of distributed computing systems including stand-alone machines, clusters, grids, clouds and edge mobile devices, all meeting a common requisite, namely: to share CPU, memory and communication capacities over a wide-area network. In [37] it was proposed to unify the view of all available computing resources (community grids, collaborative grids and cloud) by means of a grid overlay constructor. Cloud@Home is another known approach of combining cloud computing with shared resources at home [39]. In Leveraging Volunteer Computing [40], users share their own resources to be presented at cloud as virtual instances. A similar proposal for sensors and actuators can be found in [41], where a hypervisor was proposed for the abstraction and virtualization of sensors and actuators. The layer of abstraction provided by the hypervisor presents these sensors as virtual instances in cloud. The idea is interesting, and could be extended to include a broader set of edge devices. Regarding the limited resources of a sensor to be virtualized, the virtualization may be done not by slicing the sensor resources but by sharing one physical sensor among various services. In this way, each service can see the sensor as an isolated and exclusive resource, as this service is the only at a time obtaining data from that sensor.

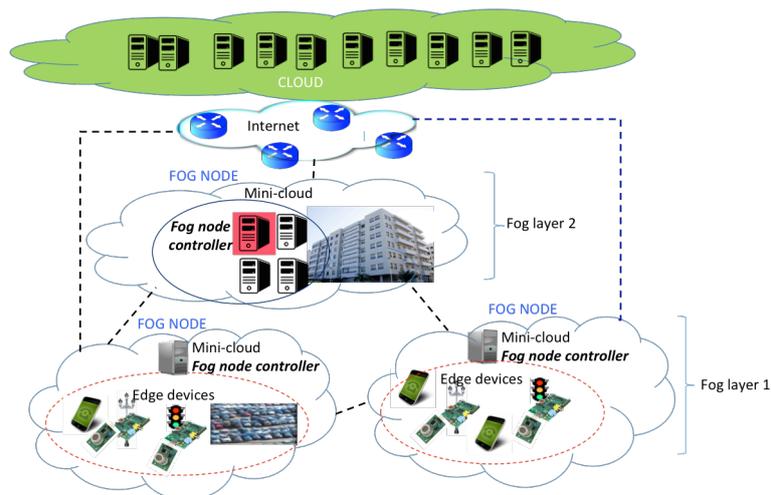

a) F2C scenario with fog nodes

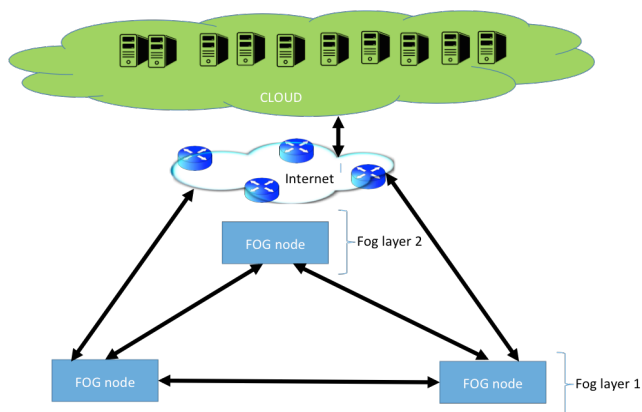

b) Logical view

**Figure 4.** *Fog node model*

### 4. Towards a formal definition for a fog node

Previous sections focused on existing contributions in fog computing relevant to the common understanding of what fog computing is and what a common definition of a fog node may become. The focus was on fog infrastructure –that is, the set of server resources which related work denotes as fog nodes- and the related edge devices connected to these nodes, whereby the work reviewed so far is more diverging than converging towards the common understanding. In this section we aim focusing on converging aspects towards common definition of a fog node, while we remain cognizant of the ongoing evolution that is blurring the differences between clouds, fogs, and edge devices as the services become more oblivious of the infrastructure used.

To start off the discussion, let us consider a hierarchical distributed Fog to Cloud (F2C) system, akin to what we have proposed in [26]. In this architecture, we assume one or more cloud layers, a few different fog layers, that also include edge devices (see Figure 1 and Figure 4). Referring back to Table 3, the proposed approach is rather similar to the ideas included in the third row that refers to "truly smart" edge devices, however, unlike any of them, we propose to integrate the edge devices' data processing capacity into the fog node definition. In this way, we can envision the orchestration and resource offloading mechanisms as the integral part of as the functions of a fog node. For this idea to work, the compute and storage capacity in edge devices should be presented in terms of virtual (abstracted) computing units. Given that, each fog node would include of two type of resources: i) one or more computing servers and ii) the aggregated capacity of the edge devices, as illustrated in Figure 3.b. In line with this concept, a fog node would not be a specific device, or a set of specific devices, but rather a logical concept, with heterogeneous type of devices as its physical infrastructure. In Figure 4 we present both the physical view and the logical view of different fog nodes in a typical F2C scenario to illustrate our idea of the logical concept of fog node. In the example illustrated in Figure 4, the mini-cloud, included in the fog node, may consist of a single server or a set of servers together building a more heterogeneous mini-cloud. At the very edge of the network, the fog node will encompass edge devices together with a mini-cloud. However, at higher levels in the F2C hierarchy, a fog node does not need to include the abstractions of edge devices, but rather only a mini-cloud. Such a layered abstraction is in fact the essence of the future joint fog-to-cloud computing architecture.

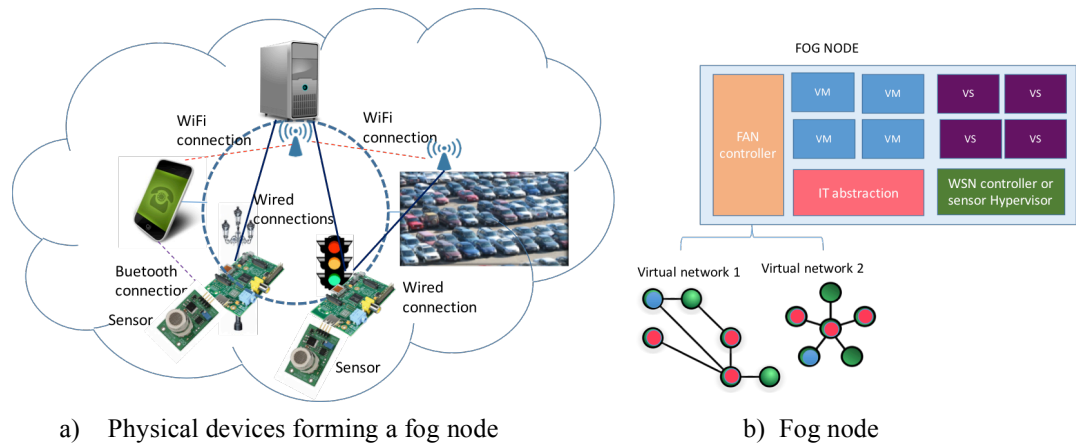

a) Physical devices forming a fog node    b) Fog node

**Figure 5.** *Fog node proposal*

How various features of edge devices can be presented as logical instances in a fog node is an open question. Also, what the computing entity in the whole system is, and its locality, where these abstractions are created and managed is open to discussion. For instance, one of the physical devices building the fog node (preferably the one with higher computing capacity) can be made responsible to deploy the abstraction, similar to the concept of cluster lead, while also granting communication between all fog layers and the cloud. In a more appropriate parlance of today's systems, this could be refereed to as the fog node controller. For instance, one of the servers, shown in Figure 4, within the mini-cloud can act as the fog node controller. The resource discovery is another open challenge, whereby various IT capacities (CPU, memory, storage) of a fog node can be presented in form of few virtualized computing unities. The same analogy applies with the sensors forming parts of a fog node, and the network connecting the fog node devices. We argue that all resources managed by a fog node should be abstracted, not only the IT resources but also the sensor and network resources. In fog-to-cloud scenario, the fog node is further responsible for presenting an abstracted and virtualized view of its resources, to higher layers, cloud (see Figure 4.b).

In sum, we believe that a fog node can be defined along the following lines:

*Fog nodes are distributed fog computing entities enabling the deployment of fog services, and formed by at least one or more physical devices with processing and sensing capabilities (e.g., computer, mobile phone, smart edge device, car, temperature sensors, etc.) All physical devices of a fog node are connected by different network technologies (wired and wireless) and aggregated and abstracted to be viewed as one single logical entity, that is the fog node, able to seamlessly execute distributed services, as it were on a single device.*

Whether this is a lasting definition, our goal is to post the question of what a fog node is in the context of a holistic, combined fog and cloud computing ecosystem, where the notion of a fog node is used to serve and present to higher layers an abstracted and virtualized view of the underlying fog resources and the networks connecting them. This is illustrated in Figure 5. Figure 5.a depicts the physical devices and the physical network forming the fog node. Figure 5.b shows a potential abstraction of the physical resources, in form of Virtual Machines (VMs), Virtual Sensors (VSs) and possible virtual networks, as seen by the cloud layers, setting all together a preliminary approach to a candidate Fog node architecture, including a FAN (Fog Area Network) controller, as well as two modules, the IT abstraction and the Wireless Sensor Network (WSN) controller. Figure 5.a also illustrates that fog node devices can be physically connected between them, using different network technologies such as 3G/4G, LTE, Ethernet, WiFi, Bluetooth, etc., but the Fog Area Controller would allow the network virtualization (virtual networks VN1 and VN2). We believe that this abstraction and the consequent integration with the cloud will not only ease the fog computing deployment, but also fundamentally change the cloud systems as we know them today, towards a more distributed and more decentralized operation, with all the qualities of the today's data center-based service provisioning.

## 5. Open Issues and Challenges on conceptualizing a fog node

In the previous sections, we made an attempt to define a fog node, emphasizing the need for and rationale behind this quest. This section focuses on open issues and challenges. Since the context of previous sections was to distinguish between capabilities of edge devices –"dumb", "smart" and "truly smart"–, we structure this section also aligned to this classification. To this end, we mainly focus on smart edge devices various processing and networking capabilities (i.e., "truly smart"), while also outlining the challenges related to "dumb" edge devices (sensors), paying particular attention on how edge devices can be virtualized, aggregated, and how to handle their mobility and information uncertainty. We finish the section with the discussion on Quality of Service, as well as security and privacy aspects.

Before going deeper into the discussion, it is important to take a moment to recognize that the scenarios envisioned are putting together a large set of highly heterogeneous resources, which creates a fundamentally complex system to be managed in a coordinated fashion. For instance, it is well known that basic computing units in the clouds are usually virtual machines. Hence, cloud management can be reduced to the management of a set of virtual machines. Moving to the edge, fog computing includes both mini-clouds as well as edge devices. Hence, in order to coordinate management between the clouds and fogs, and assuming virtualization is the strategy of choice, the compute capability of edge devices also needs to be virtualized, abstracted and aggregated. Finally, while we have emphasized the management of heterogeneous edge devices in a fog node as a great challenge, we have not yet tackled the network management issues in the context of fog computing, which further increases the complexity of the overall system analysis.

### 5.1. *Edge device virtualization*

Today, there is a plethora of heterogeneous edge devices and systems with rather different characteristics and capabilities of sensors, actuators, wearables, embedded systems, mobile phones or cars. Let us focus here on individual edge devices that have computing capabilities – CPU and memory –, hence devices with enough capacity to run some lines of code setting a service, an application, part of an application, or a function. In its basic capability, the edge device includes the hardware of the CPU and memory units, its corresponding operating system (OS), and the network interface, as illustrated in Figure 6. Figure 6.a shows the basic configuration for an edge device, including network interface, hardware and the dedicated operating system. Building on this basic system, edge device resources can be further virtualized, to optimize and extend its performance and applicability, further illustrated in Figures 6.b, 6.c and 6.d. Indeed, Figure 6.b shows an edge device whose resources are virtualized over its own OS, such as it is the case of VirtualBox [42]. A different approach is depicted in Figure 6.c where virtualization is handled using a hypervisor such as VMWare ESX-Server [43]. Figure 6.d illustrates virtualization on the operating system level, creating Linux Containers (LXC) [44] by means a known container based system Docker [45].

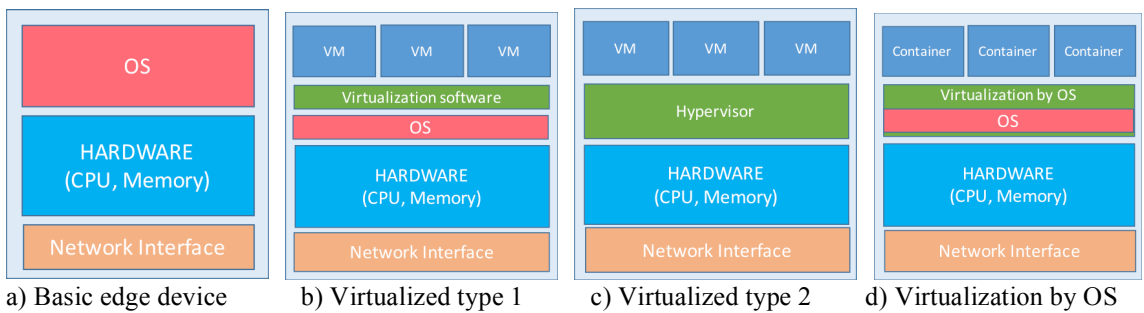

a) Basic edge device    b) Virtualized type 1    c) Virtualized type 2    d) Virtualization by OS

**Figure 6.** *Possible edge devices configuration*

Each one of the options in Figure 6 has its advantages and drawbacks. That said, we argue that, to fully integrate fog nodes with clouds, all of these scenarios need to be supported, or else, dynamic features like resource sharing cannot be implemented in an open fashion. At the same time, opens the question of the best configuration for each specific application scenario. For instance, what would be the proper configuration for a low-cost commercial board device (e.g., Arduino or Raspberry) connected to a sensor? Is the selected configuration to stay for long and what is the power consumption? If the configuration is only for short time, what is the policy triggering a configuration change? Should we guarantee that all applications use the same configuration for an edge device? Etc. By reviewing at the existing literature, Bonomi et al. in [31] suggested edge devices to be configured as either virtualized as VMs or offered as bare metal. However, other contributions, see [46], [47] and [48], use the containers to run applications in fog nodes –considered in these works as mini-clouds at the edge of the network– due to their reduced memory capacity, computing footprint, and small size. Aligned to the latter concepts, if the part of mini-clouds in the fog node is virtualized by means of containers, shall we assume that edge devices should be virtualized in the same fashion? Shall we use the same virtualization strategy for all edge devices? Indeed, edge devices capacities are much modest than a mini-cloud, hence new algorithms, methods and policies are needed to set the proper virtualization strategies for edge devices.

In sum, further research is needed to define and set policies for the best virtualization configuration when dealing with various edge devices. This is, in our understanding, one of the main challenges to be addressed when defining a fog node. With the aim of bringing some light to the challenge of edge-device virtualization, Table 4 lists some current contributions classified depending on: i) whether the fog nodes IT capabilities (as mini-clouds) are considered, and; ii) whether the IT edge devices are virtualized. References of first, second and third rows in Table 4 deal with fog nodes without including edge devices. Notice that contributions listed in the first row – *Fog-nodes virtualization not specified or bare metal* –, also include references that do not specify how fog nodes offer their resources. Furthermore, the overall set of works reviewed here also includes contributions from other research areas –vehicular clouds, grids, mobile grids, etc. We can conclude based on all work reviewed that future fog nodes should have their computing, processing and networking capabilities virtualized, and fog nodes should be able to include any type of edge device, either virtualized or not.

**Table 4.** *Mini-clouds and edge devices virtualization*

| | |
|---|---|
| **Fog node virtualization not specified or bare metal** | [1][2][4][27][29][1][28][2][32][34][29] |
| **Fog node providing virtualized resources** | [31][27][29][9] [33] |
| **Fog node offering virtualized resources as Containers** | [46][47][48] |
| **Edge device as bare metal** | [31][15] [17][3] |
| **Edge device virtualized as VMs** | [13][14][31][38][35] [39][4] |

*5.2. Abstraction and Aggregation of Edge Devices*

Recognized the need for presenting the available IT resources of a fog node as a set of available virtualized resources, such as virtual machines (VMs) or containers, we state that the whole set of virtualized resources in a fog node must encompass: i) the virtualization of the mini-cloud hardware resources, and; ii) the virtualization of the edge devices. The approach to represent a fog node as a virtual concept was earlier illustrated in Figure 3.b, where we showed that a fog node can include different virtual machines that are jointly managed. Some of these resources can be hardware resources of the mini-cloud and some other brought by an abstraction layer –represented in Figure 3.b by "Abstraction" of the edge devices–, all creating a joint topology that is hardware, software and technology agnostic. In other words, we showed that the abstraction and aggregation are the key features.

The challenges in implemented the abstraction and aggregation layer, include some of the salient features that a fog node would need to include, such as:

- *Uniform representation of edge devices*: From the fog node point of view, the heterogeneous edge devices need the same representation –in terms of characteristics, features, parameters–, which is critical to facilitating the overall management of fog computing. Every edge device would run a client software - endowed with a device manager software, yet to be defined–, that would facilitate the association with the fog node controller with the goal of that virtualization (as earlier illustrated in Figure 4).

- *Aggregation of resources*: The abstraction layer may virtualize multiple edge devices together as "aggregated". This means that abstraction is used after an aggregation process, carried out for example by clustering the resources. For instance, a gridding software (see [49][50]) deployed at the abstraction layer can aggregate and present the available resources at the edge devices as if they were from a single device. This is illustrated in Figure 7.a that depicts edge devices aggregation by a grid, also including an emulation software layer (optional). Emulation may be necessary in some cases when running an application for a type of hardware different from that provided by the edge device. For example, an x86 application is run over edge devices with ARM hardware. A different option may be to install a distributed operating system (DOS) [51][52] running in the edge devices (see Figure 7.b). Figures 7.a and 7.b depict a zoom-out view for the abstraction, both showing the advantages of aggregating the edge device computer resources with a grid or a DOS. Another example of aggregation is shown in Figure 7.c, similar to vehicular clouds (VC), where different vehicles and their IT capacities are aggregated forming a cloud, handled by specific cloud software like OpenStack [53], OpenNebula [54], etc., taking into account which cloud software would be the more suitable for architectures usually used in edge devices, such as ARM.

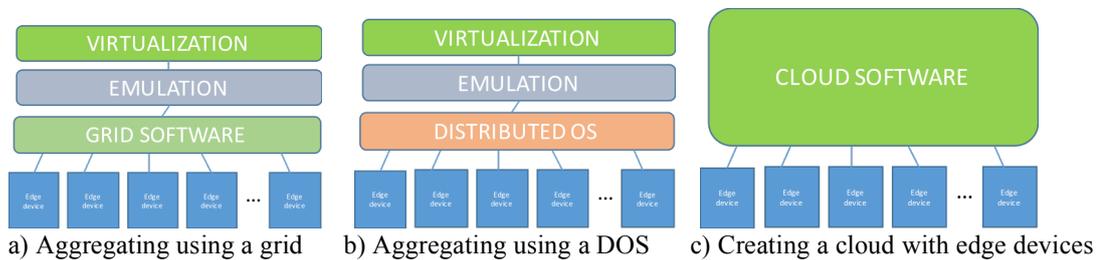

a) Aggregating using a grid    b) Aggregating using a DOS    c) Creating a cloud with edge devices

**Figure 7.** *Aggregation of edge device resources*

---

[1] Works in [27][29] argue the possibility of dealing with both type of resources, virtualized and no virtualized
[2] In [28] sensors are virtualized however there is not information about how the computer capacity of Smart Gateway acting as fog it is represented.
[3] [4] These proposals are not specifically in the fog area, but include vehicular clouds (VC), grid computing, mobile grid computing, etc.

- *Resource selection, or flexible resource aggregation*: The entire set of edge devices can be aggregated to appear as a single resource to the F2C management system. More dynamic aggregation assumes however that only a subset and a more flexible configuration of aggregated resources is possible. In other words, the grid/cloud can be set by different number of real physical resources tailored to the specific request. This flexibility can not only be more resource efficient but also contribute to a better energy management. As an example, a Raspberry Pi 2 with 4 cores Cortex A7 has a power consumption between 3,5-4 W [55], whereas an i7 consumes at least 45 W.
- *Edge devices mobility*: Some edge devices, such as a mobile phone or a car, can be on the move. In this scenario, a strategy based on volunteering was proposed in [40], where edge devices join the fog node voluntarily, leaving the system when they are not available. The reasoning behind this idea is twofold: edge devices are on the move and may leave the area of influence of the fog node; or edge devices are offline, out of battery, switched off, etc. (Next subsection is devoted to address the specific aspects related to mobility.)

In sum, the fog node definition necessitates approaches for a uniform, or standardized, view for the resources available in order to coordinate with the cloud computing management systems. The optimal aggregation strategy will depend on the type of resources and scenarios addressed, which is subject of future research.

*5.3. Mobility and Inaccuracy*

Mobility is inherent to the edge devices. In fog nodes, this makes the above discussed abstraction and aggregation even more challenging. Morevocer, a mobile edge device can be connected through a high variety of networks, be it Bluetooth, WiFi, 3G/4G/5G to a fog node. Regardless of the network technology, the fog node connectivity is generally limited by its geographical coverage area. This also means that the amount of resources through edge devices linked to a fog node is not static. By looking back at Figure 3.b, we may undoubtedly asses that the amount of VMs physically corresponding to edge devices would be dynamic.

Different proposals have addressed the mobility problem in the context of collaboration and gridding. When adopting volunteer computing (see [39], [40]), nodes voluntarily join the grid. To that end, an application installed on an edge device manages the process of joining/leaving "the grid." Often, there is a context to this decision. For example, one needs to consider the device's CPU idle cycle, or the battery life time. In [16], various solutions for mobile grid computing were considered, such as the Quality of Service, Scheduling and Resource Management, Security, Fault Tolerance, etc. One of the interesting ideas to deal with scheduling and resource management was presented in [17], where the mobility pattern of the resources was analyzed to estimate the resource availability, classifying resources into full available, partial available, and unavailable.

The cloudlet concept proposed in [11] is, to many, a synonym to the fog computing idea. In [56], a study on the impact of user mobility on cloudlet computing performance was presented, and the relationship between the user mobility patterns, the probability a device accesses a particular cloudlet and the probability of successful tasks execution was investigated. The work concludes that the user mobility affects not only cloudlet access probability but also the cloudlet computing performance. The work in [57] proposes an offloading architecture, including different heterogeneous devices, e.g., Mobile Device Clouds (MDC), cloudlets, mobile cloudlets and clouds. The estimation of the resource availability due to mobility is computed based on the history of its performance. The main outcome includes an estimation algorithm responsible for predicting the disruption factor between the device and the cloudlet as well as an estimator for the duration of the connectivity of each mobile device. Furthermore, mobile devices have an application to activate and deactivate the collaboration mode, indicating whether they openly can offer its computational power.

Specifically close to the fog computing area, the work in [58] analyses the edge device mobility from a different perspective. Fog nodes are considered static mini-clouds located at the edge of the network, whereas the edge devices are continuously moving. The most interesting contribution of this work is that based on the user movement, an event traffic application starts being processed at some fog nodes before the mobile user reaches the location predicted. Only live event processing begins at the moment the user reaches one of the fog nodes. In order to address the problem of a potential inaccurate prediction of the future location of a mobile user, authors propose to start the processing in parallel, at several locations. The location to be finally selected will be the closest to the real position when the mobile user arrives.

In the area of VANET clouds the mobility has also been addressed to a large extent and for space reasons, we mention here several references only, closest to the area we address. In [35], a specific VM migration strategy for vehicular networks was proposed, whereby different vehicles form a cloud and the mobility of one of them causes the disruption of the connection with the other vehicles. Therefore, guest VMs in this vehicle need to be migrated to either one or more of the rest of the vehicles forming the vehicular cloud, or to the roadside unit, (RSU, which are fixed stations located on the road side), or to the central cloud, depending on the resource availability. Moreover, methods are proposed to reserve part of the resources of the mobile devices to allocate migrated VMs in order to reduce the dropping rate of cloud services. A similar work can be found in [59], where authors propose to analytically model the arrival and departure of vehicles in the *Vehicular Cloud* following a Poisson distribution.

In sum, consideration of edge devices mobility is essential in important to the fog node definition. Although solutions have already been proposed, the area is wide open for research. Mobility introduces challenges in resources abstraction process. In a resource discovery process, a fog node would advertise its available capacities, including the available capacity of the underlying edge devices. In a highly dynamic scenario, however, this information may change rather frequently. It is worth noticing that the inaccuracy of the fog node management information is definitely linked to the time dimension, and holds for a specific time period. Hence, a policy to define when a fog node must perform the aggregation of resources is also required. Similar to [17], smarter policies should be investigated and applied in mobile fog computing scenarios to analyze the same.

*5.4 Network abstraction*

The fog node, including the mini-cloud located at the edge of the network as well as the edge devices in the area of the fog node, needs to be correctly managed to optimize resources utilization and services performance. The management architecture of a fog node should include challenging strategies for resource discovery, resource allocation, edge devices management. We argue that all these strategies and policies must be handled together by what we refer to as *the management plane*, from cloud to the edge, i.e., a management plane for F2C (fog-to-cloud).

In the F2C scenario envisioned, past sections first analyzed and later emphasized the need for abstraction and aggregation of the IT capabilities of the edge devices as part of the fog node resource discovery. So far, however, we have not considered the network. Recognizing the existence of the network, as the "glue" for edge devices, how can the network be abstracted and aggregated to be jointly managed with compute and storage resources? Just as there is a wide heterogeneity of edge devices, there is a wide heterogeneity in network technologies used as well, including WiFi, LTE, 3G/4G, Bluethooth, or more recently, LoRaWan [60], and SigFox [61].

Past works, like [9], [62] and [63] proposed the fog area network to be managed by means of Software Defined Networks (SDNs) or/ and Network Functions Virtualization (NFV). These proposals are aligned to the current trend of softwarization of Telecommunications [64]. In these scenarios, the fog node includes an SDN-like controller handling the programmability of the network of edge devices under the fog node control. The communication between the different fog nodes and between the fog nodes and cloud can be handled through traditional routing mechanisms –e.g., OSPF– following either a fully distributed or by a centralized management using an SDN approach as proposed in [63] where the whole network from cloud to fog is managed by SDN.

In a different set of scenarios, several contributions propose to manage the vehicular network, VANET, by means of SDN (see [65] and [66]). Authors in [65] propose to centralize the VANET network intelligence in the Road Side Unit (RSU). Then, vehicles only have to forward data packets either to other vehicles or to the RSU, based on the decisions made by the SDN controller in the RSU. The RSU is also taking the control of the overall data dissemination. The work in [66] proposes the Fog-SDN (FSDN) VANET architecture, where the fog network management is shared between the SDN controller, the RSU and the base station, all physically located at different devices. The SDN-controller sends abstract policy rules, but the final decision is taken by the RSU or the base station based on the local knowledge of its networks and resources.

Finally, the use of SDN in Wireless Sensor Networks (WSN) is worth mentioning [67]. In [67] SD-WSN (Software Defined-WSN) is proposed, enabling the separation between the data plane, formed by sensors forwarding data, and the control plane, formed by one or more controllers centralizing the network functions, such as routing or QoS. The main idea is to make the sensor network customizable by programming, which is well-aligned with the previously discussed F2C management plane objectives.

In line with a widely recognized trend in networking referred to as network softwarization, fog area networks need also to be softwarized, whereby a physical network can also be configured in terms of different virtual networks, of which the functions can run in either the fog node or the cloud. This is a highly relevant characteristic in an IoT scenario, mainly built by putting together a lot of small and heterogeneous devices forming the network. For example, while an IoT application may require a network formed by all existing sound sensors, another application may only require temperature sensors, hence claiming for a different sensor network. Towards this vision, a number of challenges need to be addressed, including the strategies for a right placement of virtual network functions.

*5.5 Quality of service*

Subsections 5.1, 5.2 and 5.3 highlight different challenges related to the abstraction of resources setting a fog node, thus including aspects related to IT, sensor and network resources. This subsection deals with QoS while the next one (see section 5.6) faces security and privacy issues.

QoS is a challenge known from cloud computing, and even though fog computing is able to address some critical aspects of QoS, such as network latency, QoS remains a challenge. The distributed nature and heterogeneity of fog devices is a large challenge. For example, due to mobility or limited battery lifetime, it is difficult to guarantee that a resource once discovered, can guarantee its presence for the duration of service lifetime. Another issue relevant to the QoS guarantees is the multi-provider environment, which in fog computing is as possible as we know it in cloud computing. The economic factors also play a role, such as whether the cellular network operator would participate in a multi-provider fog service, or the user, as owner of the end device. These and similar question require further studies.

*5.6 Security and Privacy*

Security and privacy is also a key challenge yet requiring substantial research efforts in fog computing [8]. According to Vaquero et al. [9], fog computing will inherit the security concerns coming from current virtualized environments, such as cloud computing, but augmented with the fact that fog computing is executed at the edge of the network, in a highly heterogeneous set of devices. This unquestionably makes some of the security solutions proposed for cloud computing not suitable for fog computing. One of the main security unsolved issues in fog computing is authentication at the different levels. For example, a gateway serving as fog node may be compromised or replaced by a fake one (e.g., man-in-the-middle attack). On the other hand, the fact that fog computing shifts some computational capabilities to the edge devices, drives the edge of the network to handle private, sensitive or confidential information, so highlighting issues related to privacy and trust. Thus, secure communications must be granted in order to guarantee data privacy at the edge of the network, as well as some kind of isolation mechanism when running applications (or service, or part of a service) in fog nodes.

6. Conclusions

Although early work in fog computing would have been sufficient to define a fog node as a highly virtualized platform, details were missing about the role of edge devices, as well as whether the fog nodes are to be general purpose, or defined in the context of specific applications, such as eHealth, industrial environment, Smart Cities, etc. The state of the art research identifies a fog node as a mini-cloud, located at the edge of the network, and close to the IoT devices connected to it.

In this paper, we focused on core functionalities of a fog node as well as in the accompanying opportunities and challenges towards their practical realization in the near future. We first surveyed the state of the art in technologies for fog computing, paying special attention to the contributions that analyze the role that edge devices play in the fog node definition. We then summarized and compare the concepts, lessons learned from their implementation, and show how a conceptual framework is emerging towards a unifying fog node definition. After that, we presented for the first time a logical view (Figure 4) and an architectural approach (Figure 5) about what a fog node may be. Finally, we discussed about open issues and challenges arising when the fog node has to present an abstracted and virtualized view of its physical resources (i.e., computing, sensing and networking) to higher layers in the F2C scenario.

**Acknowledgments**

This work was partially supported for UPC authors by the Spanish Ministry of Economy and Competitiveness and by the European Regional Development Fund under contract TEC2015-66220-R (MINECO/FEDER).